# Clustering Templates: A Data-Driven Framework for Dimensionality Reduction and Compression of Brain Strain in Head Impact

Junxi Liu[1], Xiaoyu Du[1], Jingyi Wu[1], Yang Wang[1], Chongmin Zhang[1], Yihan Zhang[1], Siyuan Lv[1], Li Wang[1,2,3,4], Xianghao Zhan[5], Zhou Zhou[6], Yuzhe Liu*[1,2,3,4]

**Corresponding Author**

Yuzhe Liu

Email: yuzheliu@buaa.edu.cn

**Affiliations**

[1] School of Biological Science and Medical Engineering, Beihang University, Beijing, 100191, China

[2] Key Laboratory of Biomechanics and Mechanobiology (Beihang University), Ministry of Education, Beijing, 100191, China

[3] Key Laboratory of Innovation and Transformation of Advanced Medical Devices, Ministry of Industry and Information Technology, Beijing, 100191, China

[4] National Medical Innovation Platform for Industry-Education Integration in Advanced Medical Devices (Interdiscipline of Medicine and Engineering), Beijing, 100191, China

[5] Department of Bioengineering, Stanford University, Stanford, CA, 94305, USA

[6] Division of Neuronic Engineering, KTH Royal Institute of Technology, 14152 Stockholm, Sweden

**ORCIDs**

Junxi Liu: 0009-0002-0375-6957

Xianghao Zhan: 0000-0002-7058-2603

Zhou Zhou: 0000-0002-3910-0418

Yuzhe Liu: 0000-0002-9451-0620

# Abstract

**Purpose:** Finite element (FE) models are indispensable for studying the biomechanics of traumatic brain injury (TBI), and modern head FE models typically comprise thousands to millions of elements. The resulting brain strain is therefore high-dimensional, and when combined with dense temporal sampling and large numbers of head impact cases, this produces datasets of substantial data volume. Accordingly, this study investigates the effective dimensionality required to accurately represent brain strain and proposes a data-driven framework to reduce the dimensionality of FE simulation data while maintaining high spatiotemporal fidelity. **Methods:** We developed a series of "clustering templates" by applying K-means clustering to feature vectors derived from the maximum principal strain (MPS) time-history curves of 118,840 brain elements over 3,693 impact simulations using the THUMS head model. The performance of these templates, with cluster counts (*K*) ranging from 8 to 8192, was rigorously evaluated by quantifying the fidelity of data reconstruction. **Results:** The clustering templates demonstrated high-fidelity reconstruction of the original FE data. A template with *K* = 512 achieved a compression ratio of over 200:1 while maintaining a low mean element-wise peak MPS root mean square error (RMSE) of 0.027, and a template with *K* = 4096 (30:1 compression ratio) achieved an RMSE of 0.016. **Conclusion:** The proposed clustering templates provide an effective and automated method for the dimensionality reduction and compression of large-scale brain strain data. This framework mitigates critical data management bottlenecks and enables the development of subsequent data analysis and machine learning models, thereby accelerating research in TBI biomechanics.

# Introduction

Traumatic Brain Injury (TBI) is a significant global health threat. It can be caused by various events, including falls, traffic accidents, collisions in sports, and blast exposures [1–3]. A global study indicated that over 20 million new cases of TBI were reported in 2021 alone [4]. TBI is a leading cause of mortality worldwide and can lead to severe neurological sequelae, such as cognitive impairment and disability, which profoundly impact a patient's employability, independent living, and overall quality of life [5–7]. A major challenge in TBI management is that initial computed tomography (CT) scans may appear normal, yet patients can later develop intracerebral hemorrhage and secondary neurological deterioration, making early diagnosis and treatment exceptionally difficult [8].

To deepen our understanding of TBI biomechanics and to provide quantitative metrics for early diagnosis, Finite Element (FE) analysis has become an indispensable research tool [9]. By applying kinematic data to FE models of the human head— such as the KTH FE model [10], the Global Human Body Models Consortium (GHBMC) models, or the Total Human Model for Safety (THUMS) [11] and Worcester head injury model (WHIM) [12]— we can obtain the spatiotemporal distribution of key metrics like Maximum Principal Strain (MPS) that are crucial for investigating the mechanisms of TBI and developing methods for evaluating TBI risk [13].

As shown in **Table 1**, the demand for higher biofidelity of anatomy has led to increasingly fine meshes in head FE models, characterized by a growing number of smaller elements. For models with high mesh resolution designed to capture the complex geometry of the human head (e.g., sulci and gyri), this level of detail is necessary due to the small volume of these structures [14, 15]. However, for models that do not retain such detailed sulcal-gyral anatomy, such fine meshes often introduce redundancy in further analyses. Furthermore, the high-dimensional data generated by fine-meshed FE simulations presents significant challenges for data storage, analysis, and further machine learning applications.

**Table 1. Number of brain elements in different versions of various head FE models.**

| Model Name | Year | Number of elements |
|---|---|---|
| KTH | 2002 | 18416 [16] |
| | 2021 | ~4,940,000 [17] |
| WSUHIM/GHBMC (Head) | 1993 | 7,351 [18] |
| | 1995 | 22,995 [19] |
| | 2001 | 314,500 [20] |
| | 2013 | 270,552 [21] |
| | 2019 | 82,083 [22] |
| | 2022 | 246,829 [23] |
| THUMS (Head) | 2007 | 49,700 [11] |
| | 2018 | 62,048 [24] |
| | 2021 | 61,345 [25] |
| WHIM | 2014 | 115,228 [26] |
| | 2019 | 143,300 [12] |
| | 2020 | 202,800 [27] |
| | 2022 | 202,800 [28] |

One application of FE simulation is the deep learning-based brain strain computing models, which can rapidly calculate the brain strain according to the head kinematics. Wu et al. proposed using a Convolutional Neural Network (CNN) to predict MPS from head kinematics, achieving a more than 36,000-fold speed-up on a standard computer, compared to the conventional FE simulation [29]. Their research has expanded the application of CNNs to include element-wise MPS distribution prediction [30], regional strain prediction, and integration with helmet safety evaluations [31]. Meanwhile, Zhan et al. [32] independently developed a fully connected neural network to predict the spatial distribution of peak MPS, with later work incorporating transfer learning [33] and domain adaptation [34] to reduce errors arising from differences among head impact datasets. Training such models typically requires large numbers of head impact cases due to the high dimensionality of MPS outputs, yet acquiring sufficiently large and diverse experimental datasets remains challenging [35, 36]. To alleviate this data demand, Wu et al. [29, 30] generated synthetic head kinematics by reorienting measured kinematic axes. However, because both the human head–neck system and brain tissue exhibit anisotropic properties, such synthetic kinematics may not fully capture the characteristics of real-world head impacts. To further reduce MPS data redundancy and thus alleviate the data needs, Zhan et al. [37] applied principal component analysis (PCA) to mitigate collinearity in MPS data, thereby simplifying the prediction task and reducing training data requirements; however, this approach focused only on spatial variability and did not assess reconstruction fidelity. Nonlinear manifold-learning approaches, such as Grassmannian diffusion maps, have also been applied to high-dimensional computational head model outputs, particularly for surrogate-based uncertainty quantification of full-field strain responses [38].

Another application of FE simulation is TBI risk modeling, in which brain strain data are used as inputs to predict injury risk. Cai et al. [39] utilized voxel-wise or region-wise strain fields for machine learning–based concussion classification, highlighting the predictive value of high-dimensional strain representations. Wu et al. [40] proposed a network-based response feature matrix that encodes distributed strain information, including maximum principal strain (MPS) and white matter fiber strain, into a connectivity-informed representation, demonstrating improved concussion classification compared with scalar metrics. However, these approaches rely on hand-crafted features or direct use of strain fields, resulting in rapidly growing feature dimensionality and the associated "curse of dimensionality." This phenomenon has been widely observed in high-dimensional medical imaging analysis and bioinformatics [41, 42]. It leads to unstable model training, excessive memory demands, and an increased risk of overfitting when clinical datasets are limited. Specifically, when the sample size is much smaller than the feature dimensionality (e.g., limited clinical cases vs. high-dimensional FE data), models tend to memorize the training examples rather than learning generalizable injury patterns [43]. Moreover, the high dimensionality of FE outputs fundamentally limits interpretability, even for models that are intrinsically transparent (e.g., logistic regression). While such models allow for direct inspection of feature weights in low-dimensional settings, the sheer volume of parameters here renders manual verification cognitively intractable, making the decision-making process of models a "black box" [41].

Furthermore, most studies typically represent brain strain in a head impact using single scalar metrics, such as MPS95 and the Cumulative Strain Damage Measure (CSDM) [44–52]. MPS95 is the peak of the 95th percentile MPS time-history curve across all elements at each time point, and CSDM is the ratio of brain elements experiencing MPS values higher than a specific threshold during the entire impact. While computationally convenient, these manual feature extraction methods risk losing critical information about the brain's mechanical response. For instance, regarding the 95th percentile MPS time-history curve, Rooks et al. [53] found that using MPS95 as a brain strain metric resulted in the loss of impact direction information and, consequently, the loss of regional strain characteristics. This curve only captures the magnitude of high strains within the brain at each moment but discards information about the spatial location of these high-strain regions. For a single impact event, the MPS trends can vary substantially across different brain regions [54]. This spatial information is potentially crucial for predicting injuries like TBI, as certain brain regions may be more vulnerable and sustain damage even at lower strain levels. In addition, many impacts may be very similar on a certain metric, but vastly different in modalities other than this metric [55]. Because these manually-defined metrics rely on human intuition rather than the data itself, they may introduce subjective biases [41]. Therefore, this study aims to develop an automated method that achieves efficient dimensionality reduction while maximally preserving the original spatiotemporal information, thereby addressing the shortcomings of existing strategies.

To solve these problems, this study proposes a general reduced representation of head FE data, termed a **clustering template**, developed through a data-driven dimensionality reduction framework. The fundamental hypothesis of this work is that many brain elements exhibit similar mechanical responses, that is, they follow similar MPS time-history curves when the head is subjected to impacts from various directions. By leveraging this redundancy, we employ an unsupervised clustering method to group the elements of the THUMS model into multiple **super-elements**. Since the THUMS elements within each super-element share similar MPS time-history curves, we can use the average MPS time-history of the super-element to represent each of its constituent elements, thereby achieving data dimensionality reduction and compression. This method learns from a large corpus of THUMS FE data to abstract the high-fidelity THUMS mesh into a "super-mesh" on a "clustering template." Without introducing human assertions or biases, this approach efficiently simplifies the original high-resolution spatial output of the FE model into a general spatial-reduced representation, which may help alleviate the dimensional bottlenecks for downstream machine learning applications, such as high-dimensional surrogate modeling and injury classification. By comparing the THUMS data reconstructed from our clustering templates with the original data, we demonstrate that this method incurs only small information loss (quantified by high CCC and low RMSE), paving the way for subsequent data analysis and model training that rely on FE data.

# Materials and Methods

### Data Preparation

We generated a synthetic dataset comprising 1,998 unique impact simulations by systematically varying the impact magnitude and direction within a validated Hybrid III head-neck and impactor model system [56]. These simulations were designed to cover a wide range of conditions, ensuring comprehensive coverage of potential impact scenarios. In addition to this synthetic dataset, we integrated several real-world kinematic datasets, including those from college football games (n = 302), high school football games (n = 602) and mixed martial arts (MMA) bouts (n = 463) collected with the Stanford instrumented mouthguard [3, 48, 57–61], the National Football League (NFL) (n = 53) [62], and the National Association for Stock Car Auto Racing (NASCAR) (n = 275) [63]. The final dataset consisted of both synthetic and real-world impacts, with synthetic simulations accounting for approximately half of the total cases. Across the 3,693 impacts, the observed peak linear acceleration, angular velocity, and angular acceleration ranged from 5.20 to 220.71 m/s², 3.64 to 91.78 rad/s, and 480.16 to 37,823.68 rad/s², respectively. Their distributions and directional characteristics are shown in Supplementary Fig. S2.

All kinematic data were sampled at 1 kHz, with durations ranging from 18 ms to 200 ms and an average duration of 95 ms. The final consolidated dataset, comprising $N_{\text{impact}}$ = 3,693 impacts, was used to drive the THUMS head model in the LS-DYNA solver (**Fig. 1a**). The THUMS AM50 Pedestrian Model Version 4.02 [11] was selected as our target model, which contains $N_{\text{elem}}$ = 118,840 brain elements. The model employs a smoothed brain surface without gyral or sulcal geometry, and the element set encompasses the cerebrum, cerebellum, and brainstem, all of which were included in the analysis. These tissues are characterized using the linear viscoelastic material law. For each impact, the simulation generated an MPS time-history curve for every THUMS element, resulting in an output tensor of shape $T \times N_{\text{elem}}$, where $T$ denotes the total number of output frames. If a simulation terminated prematurely due to numerical errors (e.g., negative element volume), the resulting MPS data were truncated at the last successfully computed time step and included in the dataset. Among the 3,693 simulations, 36 (approximately 0.97%) terminated prematurely due to such numerical errors.

## Feature Extraction

Directly concatenating the full MPS time series for each THUMS element would create an excessively large feature space, making subsequent computations intractable. To create a compact yet representative feature vector for each element, we implemented the following extraction strategy. For each of the $N_{\text{elem}}$ brain elements, we iterated through all $N_{\text{impact}}$ (3,693) impact events. Within each impact, we first identified the global peak time, $t_m$, by finding the maximum of the 95th percentile MPS curve (**Fig. 1b**). This curve was generated by taking the 95th percentile MPS value across all THUMS elements at each time step. We then extracted the MPS values at four predefined time points centered around $t_{\text{m}}$: $t_{\text{m}}$ - 6ms, $t_{\text{m}}$ - 2ms, $t_{\text{m}}$ + 2ms, and $t_{\text{m}}$ + 6ms (**Fig. 1c**). These time offsets were selected to provide a compact and computationally efficient sampling of the local ascending and descending portions of the strain response surrounding the global high-strain phase. We excluded the peak time $t_{\text{m}}$ itself to mitigate the influence of potential numerical artifacts (e.g., MPS values exceeding 1.0) in the simulation, which were most likely to occur at $t_{\text{m}}$. By focusing on the surrounding time points, this strategy effectively bypassed peak-related numerical artifacts while still accurately characterizing the overall temporal profile of the strain response. To further reduce the impact of numerical errors, these feature values were first clipped at an upper limit of 1.0 and then subjected to a log(x+1) transformation. Subsequently, for each brain element, we concatenated the features generated from all $N_{impact}$ (**Fig. 1d**). This process yielded a feature vector for each element with a dimensionality of $d = 4 \times N_{impact}$ (i.e., $d = 4 \times 3{,}693 = 14{,}772$). This high-dimensional vector served as a signature of the biomechanical response of a single THUMS element across all impact patterns in the dataset.

## Normalization and Clustering

To ensure that each feature contributed equally to the distance calculations in the K-means algorithm, we applied Z-score normalization to all features after the log(x+1) transformation. Specifically, from each feature value, we subtracted the mean of that feature across all brain elements and then divided by its standard deviation. This resulted in a normalized feature distribution with a mean of 0 and a standard deviation of 1. Following standardization, the K-means clustering algorithm was employed to group the brain elements based on their feature vectors. The K-means algorithm partitions the elements into $K$ clusters by iteratively performing two steps: (1) assigning each element to the cluster whose centroid is nearest in Euclidean distance, and (2) updating each centroid as the mean of all elements assigned to it. These steps are repeated until cluster assignments converge [64, 65]. The number of clusters, $K$, determines the level of granularity at which the cluster centroids represent their members (i.e., the THUMS elements within the cluster). To accommodate the varying data requirements of downstream applications, such as machine learning inputs and efficient data storage, we performed K-means clustering for a range of $K$ values (8, 16, 32, 64, ..., 8192) (**Fig. 1e**). The cluster centroids were initialized using the standard Forgy method [66], which involves randomly selecting $K$ THUMS elements as the initial centroids.

## Projection between THUMS and Template

K-means clustering provides a data-driven method for simplifying the high-resolution THUMS mesh. By grouping elements with similar biomechanical response signatures, we effectively created a series of coarsened **template super-meshes**. Each template comprises K response-based groups, or **template super-elements**. Here, the term “super-mesh” refers to a response-based reduced representation and does not imply spatial connectivity; the constituent THUMS elements of a template super-element are not necessarily spatially contiguous. To facilitate rapid projection between the original THUMS mesh and the template super-meshes, we constructed two projection matrices:

1. **THUMS-to-Template Matrix**: An $N_{elem} \times K$ matrix was used to project data from the THUMS mesh space to the template space. For any given data in the THUMS mesh space, multiplication by this matrix performs element-wise averaging over all THUMS elements that constitute each of the $K$ template super-elements. This operation transforms data from a shape of $* \times N_{elem}$ in THUMS space to $* \times K$ in the template space. Before multiplying, we recommend clipping the MPS results at an upper limit of 1.0 to avoid numerical artifacts.

2. **Template-to-THUMS Matrix**: A $K \times N_{elem}$ matrix was used to project data from the template space back to the THUMS mesh space. For any given data in the template space, a dot product with this matrix assigns the value of a super-element to all its constituent THUMS elements. This operation transforms data from a shape of $* \times K$ in the template space to $* \times N_{elem}$ in the THUMS space.

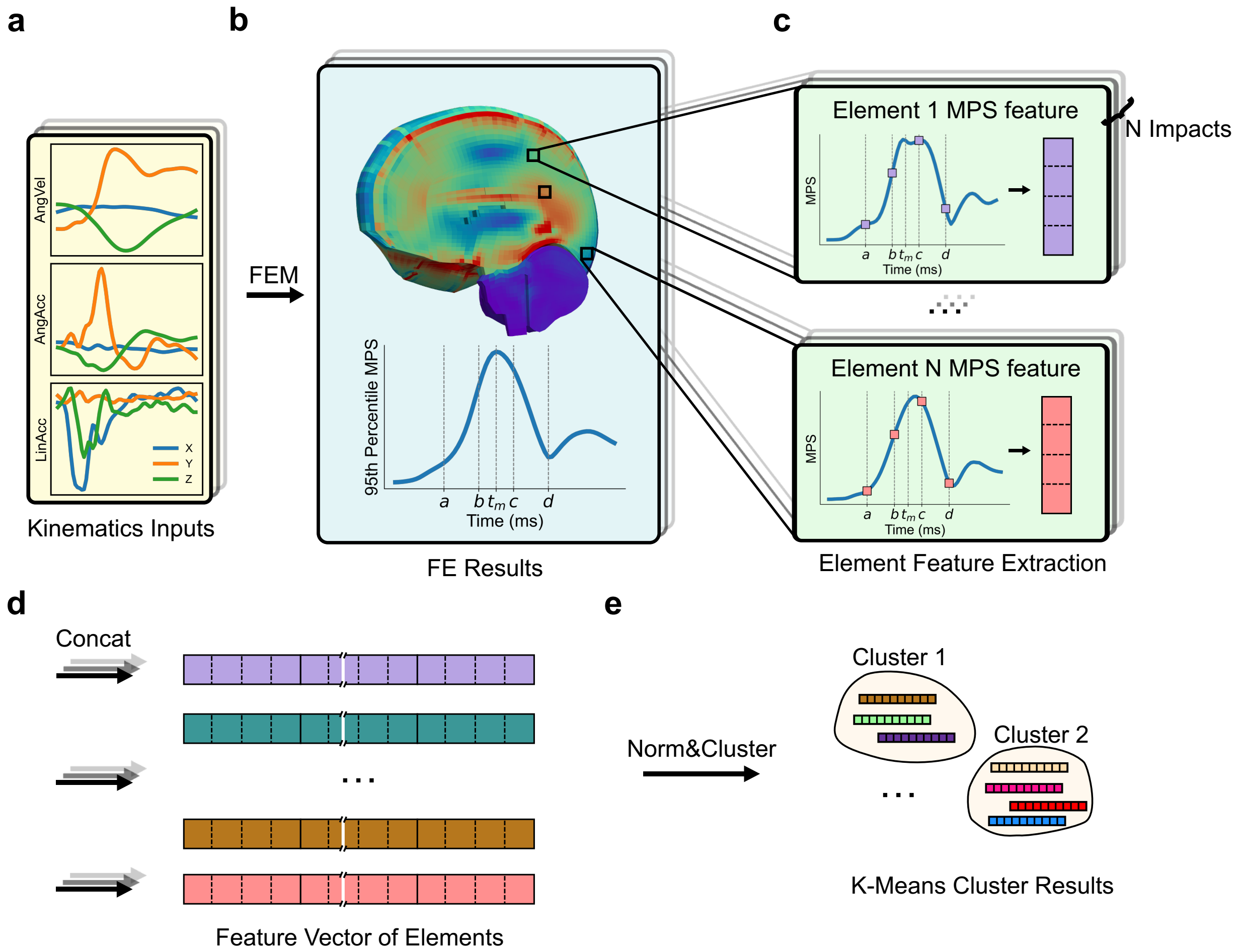

**Fig. 1 Methodological workflow for creating the clustering template. (a)** Kinematic data for each impact serve as inputs to the FE model. **(b,c)** FE simulation results and the feature extraction strategy. $t_m$ denotes the time of the peak of the whole-brain 95th percentile MPS curve. Four features are extracted at time points a, b, c, and d, corresponding to $t_m$ - 6ms, $t_m$ - 2ms, $t_m$ + 2ms, and $t_m$ + 6ms, respectively. **(d)** Concatenation of features from all impacts for each THUMS element. **(e)** Standardization of the feature matrix followed by K-means clustering.

# Results

To evaluate the extent to which the template super-meshes represent the original THUMS mesh, we performed a reconstruction analysis. We used the THUMS-to-template matrix to project the MPS data from the THUMS mesh space into the template super-mesh space. Then, we used the template-to-THUMS matrix to reconstruct the data back into the THUMS mesh space. By comparing the original and reconstructed MPS data, we quantified the information loss introduced by the template-based compression. To comprehensively assess the performance of the templates, we systematically compared the reconstructed data with the original FE data across four aspects: overall spatiotemporal distribution, element-wise peak values, time-history curve morphology, and quantitative error metrics across the entire dataset.

### Spatiotemporal Reconstruction Fidelity

We selected a representative impact case and visually compared its original MPS data with the data reconstructed from templates with different $K$ values, as shown in **Fig. 2**. Each row corresponds to a specific $K$ value (with the bottom row showing the original THUMS data), displaying the MPS distribution at various time points during the impact. The images reveal that as $K$ increases, the reconstructed MPS distribution progressively converges to the original THUMS data. At a sufficiently large $K$ (e.g., $K$ = 8192), the reconstructed MPS data become virtually indistinguishable from the original, indicating that our templates can effectively preserve the original spatiotemporal information.

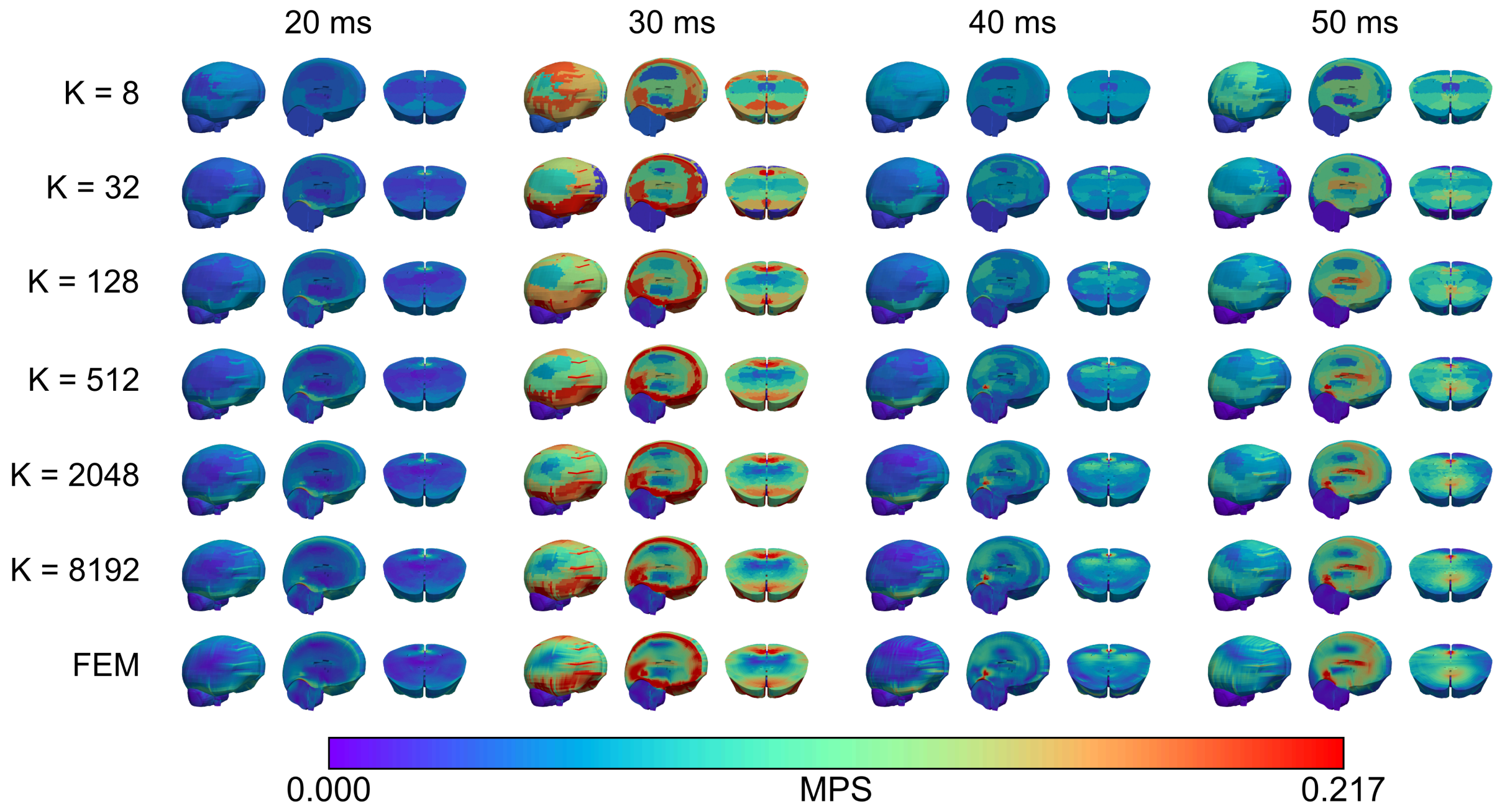

**Fig. 2 Visualization of spatiotemporal MPS reconstruction for a representative impact case using templates with varying *K* values.** The 95th percentile peak MPS of this impact is near the median of the entire dataset, ensuring its representativeness.

### Element Peak Strain Reconstruction Fidelity

The peak MPS of each element is a commonly used metric for predicting brain injuries such as concussion. Therefore, we also evaluated the reconstruction fidelity in this dimension. **Fig. 3** visualizes the absolute error between the original and reconstructed peak MPS for each element across impacts of varying intensity. The heatmaps demonstrate that for the vast majority of brain elements, the reconstruction error remained low, as reflected by the localized and progressively diminishing error distributions. While larger errors are initially observed in localized regions at lower *K* values, these error "hotspots" rapidly shrink and fade as *K* increases. This demonstrates that, at the level of element-wise peak strain, the template super-elements serve as faithful representatives of their constituent THUMS elements.

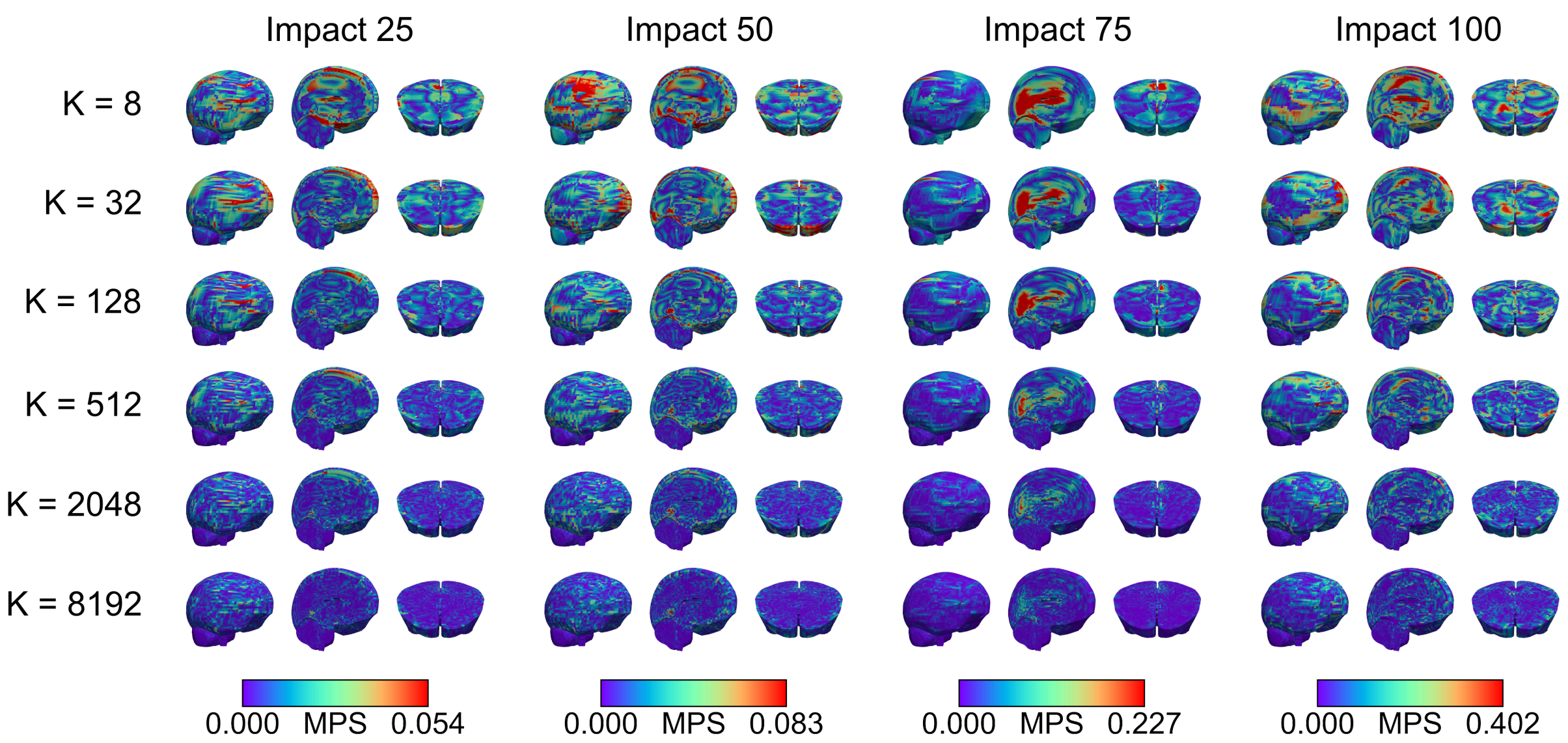

**Fig. 3 Reconstruction error of element-wise peak MPS for impacts of varying intensity across different *K*-value templates.** The intensities of Impact 25, Impact 50, Impact 75, and Impact 100 (based on 95th percentile peak MPS) are near the lower quartile, median, upper quartile, and maximum of the dataset, respectively.

### Strain Time-History Representing Capability

To assess how well the template super-elements represent the dynamic response of their constituent THUMS elements, we selected representative super-elements and plotted their time-history curves alongside the curves of all the individual THUMS elements they contain for a given impact (**Fig. 4**). The plots show strong agreement between the individual element curves and the representative super-element curve. As *K* increases, the dispersion of the individual element curves around the central super-element curve diminishes, eventually converging to a near-perfect match. This indicates that the super-elements effectively capture the collective temporal dynamics of the elements within them.

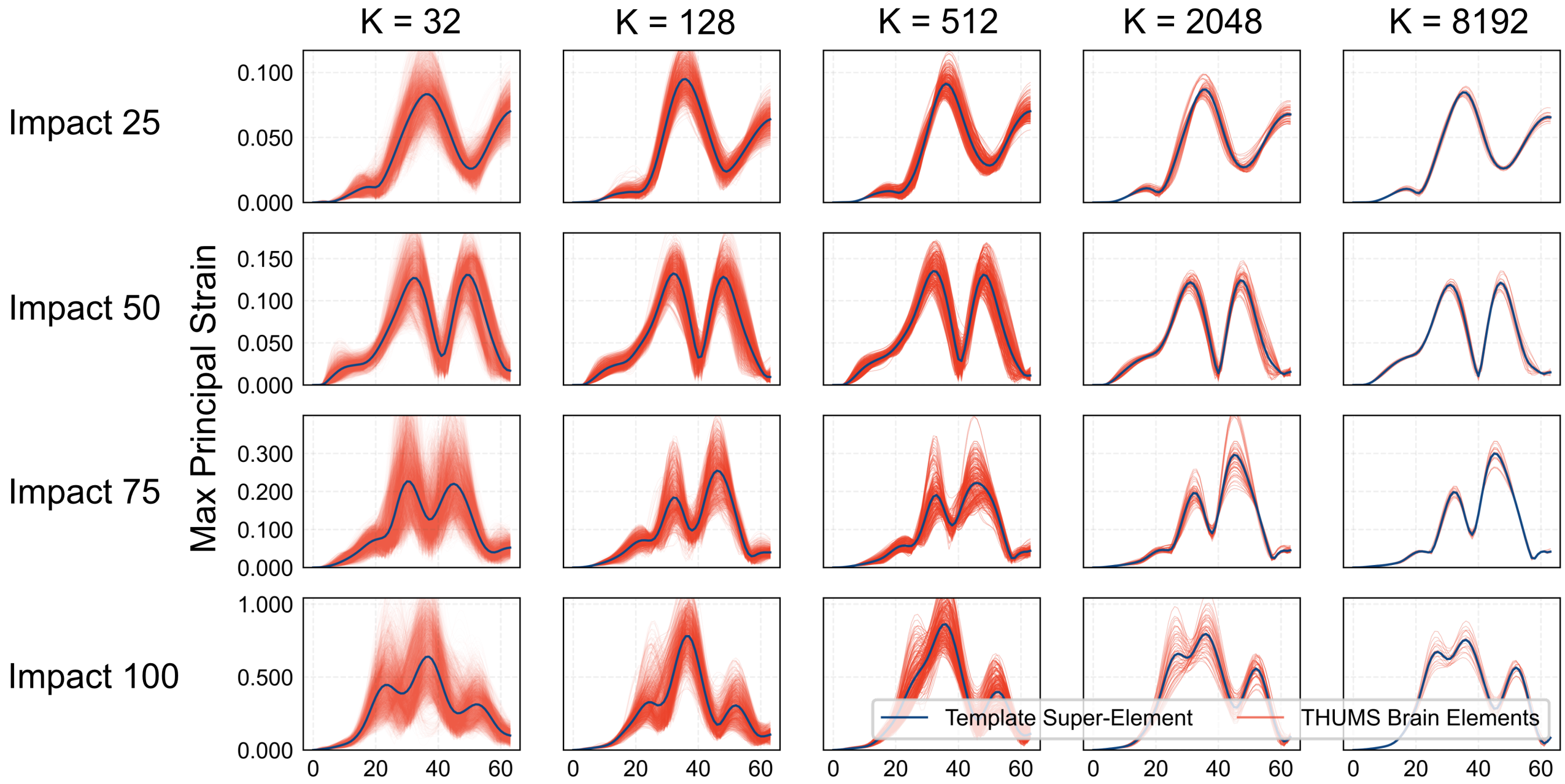

**Fig. 4 Comparison of the MPS time-history curve of a randomly selected super-element with the curves of its constituent THUMS elements.** The comparison is shown for impacts of varying intensity and templates with different *K* values.

### Quantitative Error Metrics

We applied several error metrics, including Root Mean Square Error (RMSE), Mean Absolute Error (MAE), and Concordance Correlation Coefficient (CCC) [67], to quantitatively compare the reconstructed MPS data with the original data across the entire dataset. Prior to all quantitative analyses, MPS values were clipped at an upper limit of 1.0 to mitigate the influence of numerical artifacts, consistent with the preprocessing pipeline described in the Methods. **Fig. 5b** shows the distribution of element-wise peak MPS absolute error for a representative impact case, whose kinematics are shown in **Fig. 5a**. **Fig. 5c** illustrates the distribution of the peak MPS RMSE for each impact event across the dataset. **Table 2** provides a comprehensive summary of the reconstruction performance for templates with different *K* values. All error metrics consistently improve as *K* increases, confirming that the reconstruction fidelity systematically rises with the number of clusters. At the coarsest resolution ($K = 8$), the model exhibited a performance with a Spatial RMSE of 0.062 and a Spatiotemporal CCC of 0.829. However, as the number of clusters increased to $K = 512$, the reconstruction fidelity already achieved a highly robust level: the Spatial RMSE dropped significantly to 0.027, while the CCC improved to 0.967, indicating that a moderate number of clusters can capture the majority of the signal information. At the finest resolution ($K = 8192$), the errors were further minimized to a negligible level, yielding a Spatial RMSE of just 0.014 and a near-perfect CCC of 0.992. These quantitative results affirm that our clustering template super-meshes can preserve the FE-derived MPS response signatures from the original THUMS mesh with high fidelity, as evidenced by the high correlation and low reconstruction errors.

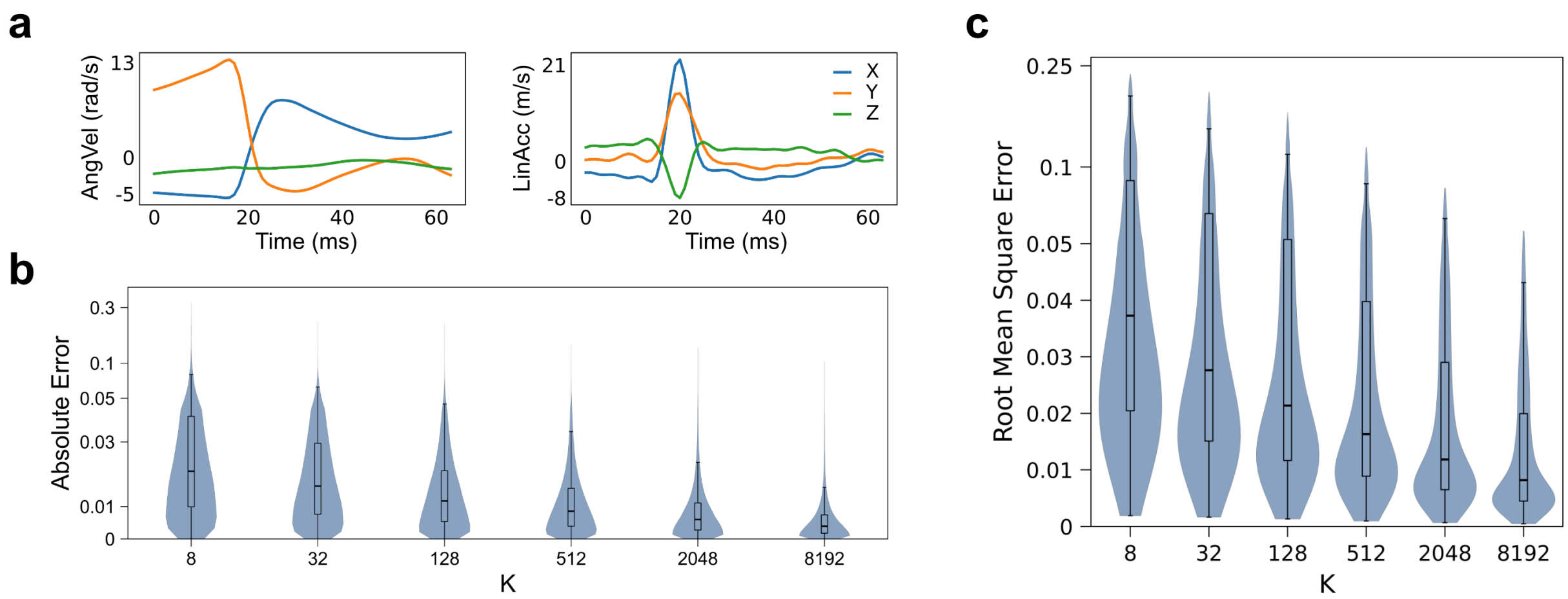

**Fig. 5 Quantitative analysis of reconstruction error. (a)** The input head kinematics for Impact 50. **(b)** Distribution of absolute error in element-wise peak MPS reconstruction for Impact 50 across different *K*-value templates. See Supplementary Material for additional impact cases. **(c)** Distribution of the root mean square error (RMSE) of peak MPS reconstruction for each impact event in the dataset ($N_{\text{impact}}$ = 3,693).

**Table 2. Reconstruction performance metrics for templates with varying *K* values.** "Spatiotemporal" refers to the reconstruction of the entire MPS time-history data, while "Spatial" refers to the reconstruction of element-wise peak MPS.

| *K* | Spatiotemporal (Element time-history curve) | | | Spatial (Element Peak) | |
|---|---|---|---|---|---|
| | RMSE | MAE | CCC | RMSE | MAE |
| 8 | 0.031 | 0.019 | 0.829 | 0.062 | 0.044 |
| 16 | 0.026 | 0.015 | 0.889 | 0.050 | 0.036 |
| 32 | 0.023 | 0.014 | 0.912 | 0.045 | 0.033 |
| 64 | 0.021 | 0.013 | 0.928 | 0.041 | 0.030 |
| 128 | 0.019 | 0.011 | 0.944 | 0.036 | 0.025 |
| 256 | 0.017 | 0.010 | 0.957 | 0.031 | 0.022 |
| 512 | 0.015 | 0.008 | 0.967 | 0.027 | 0.019 |
| 1024 | 0.013 | 0.007 | 0.976 | 0.023 | 0.016 |
| 2048 | 0.011 | 0.006 | 0.983 | 0.020 | 0.014 |
| 4096 | 0.009 | 0.005 | 0.988 | 0.016 | 0.011 |
| 8192 | 0.007 | 0.004 | 0.992 | 0.014 | 0.009 |

To further assess the sensitivity of the reconstruction performance to random initialization of the K-means algorithm, we performed an additional repeated-initialization analysis for a representative template resolution, *K* = 2048. Specifically, the clustering procedure was repeated three times using different random initial centroids, and each resulting template was evaluated using the same projection–reconstruction procedure. Consistency was assessed based on the resulting reconstruction errors rather than on exact cluster-label agreement. The reconstruction errors were highly similar across the three repetitions, with spatiotemporal RMSE values of 0.01079, 0.01084, and 0.01080, respectively. These results indicate that, for this representative template resolution, random initialization had only a limited influence on reconstruction error.

# Discussion

In this study, we introduced an unsupervised clustering method to address the challenges posed by the large number of elements in head FE models, namely the difficulties in data storage and downstream deep learning model training. We developed a series of **clustering templates** to perform dimensionality reduction, creating general coarser-grained representations of the brain mesh. Our findings demonstrate that data projected into the template "super-mesh" space can be reconstructed back to the original THUMS mesh space with low reconstruction error, as quantified by RMSE, MAE, and CCC. This data-driven approach offers significant advantages over previous methods based on manual feature extraction [29, 32].

Prior deep learning efforts to process spatiotemporal MPS data have often relied on manual "feature engineering," extracting hand-crafted metrics such as peak values or pulse widths. However, these methods are based on human intuition rather than intrinsic data patterns and risk discarding key information relevant to injury. Moreover, because handcrafted metrics usually collapse the distributed strain field into a limited number of predefined scalar or regional quantities, the original spatiotemporal MPS field generally cannot be reconstructed from them. Therefore, although such metrics are compact and useful for specific analyses, they may discard spatial distribution, regional heterogeneity, and time-history morphology that could be relevant to downstream tasks. Furthermore, high-dimensional spatiotemporal data places a significant burden on storage systems, increasing costs and creating bottlenecks that hinder data sharing among research groups. While methods like Principal Component Analysis (PCA) have been explored for dimensionality reduction of FE data [37], PCA represents strain fields as linear combinations of global basis vectors. The learned principal components are dataset-dependent and may need to be recomputed or revalidated when the impact dataset changes. In addition, PCA coefficients do not directly correspond to spatially defined brain regions or physically interpretable strain quantities. Besides, as a linear technique, PCA may fail to capture the complex, non-linear relationships inherent in the data depending on the application.

Beyond linear approaches such as PCA, nonlinear manifold-learning methods have also shown promise for handling high-dimensional outputs from computational head models. For example, Grassmannian diffusion maps have been used to construct low-dimensional representations of full-field brain strain responses and, when combined with Gaussian process regression and geometric harmonics, to support surrogate-based uncertainty quantification of head model predictions [38]. This line of work further suggests that high-dimensional brain strain fields may contain an intrinsic lower-dimensional structure that can be exploited for efficient computation and analysis. The clustering-template framework proposed here addresses this opportunity from a complementary perspective. Instead of learning a latent manifold for input–output surrogate prediction, our approach constructs a reusable mesh-space representation by grouping THUMS elements with similar MPS time-history response signatures into template super-elements. This representation enables direct projection between the original THUMS mesh and the lower-dimensional template space, making it suitable for data compression, storage, reconstruction, and preprocessing for downstream machine-learning tasks. In future work, clustering templates may also be combined with nonlinear manifold-learning or surrogate-modeling approaches, for example by first reducing element-level FE outputs into template space before learning lower-dimensional response manifolds or input–output mappings.

By projecting data from the THUMS mesh space to our template space, we can reduce the data dimensionality from $N_{elem}$ to $K$. By selecting an appropriate $K$ based on the needs of a downstream application, data volume can be compressed by factors ranging from tens to thousands. This transforms high-dimensional, intractable data into a manageable, lower-dimensional representation in a data-driven manner. Depending on the chosen scale of $K$, this reduced representation serves dual approximation objectives. At higher resolutions (e.g., $K$ = 4096), the template is designed to approximate the full, element-wise strain time-history traces with high fidelity. Conversely, at lower resolutions (e.g., $K$ = 16), local high-frequency details are filtered, and the template transitions to representing the overall summary behavior of the brain, capturing macroscale regional co-deformation trends. This parameterizable design allows researchers to dynamically balance the retention of full temporal traces against the abstraction of broad biomechanical patterns.

This method may facilitate the development of downstream deep learning models, particularly when labeled samples are limited, because the template-space data remain FE-derived MPS quantities while reducing the output dimensionality. For instance, training a model to predict TBI occurrence from MPS data is currently challenging due to the difficulty in defining input features and the scarcity of labeled clinical data. However, after simplifying the spatial dimension with our clustering templates, the smaller input feature space reduces the amount of data required for training. For templates with smaller $K$ values, we also reduce the extremely high-dimensional FE output to a scale that is amenable to human interpretation, facilitating a more intuitive understanding of brain strain data. This will greatly promote further research on head FE data, such as investigating the strain sensitivity of different brain regions or exploring the correlations in their mechanical responses.

A major potential contribution of this work is the reduction of storage costs and the promotion of data sharing. Previously, the sheer volume of FE data has hampered storage and collaboration. Our dimensionality reduction method allows researchers to compress data according to their specific needs, eliminating redundancy and thereby improving storage efficiency and transfer speeds. For example, the raw spatiotemporal MPS data for a single impact (118,840 elements × 100 time steps × 8 bytes/double) occupies approximately 90 MB. Using a $K$ = 512 template, the data can be compressed to less than 0.5% of its original size (512 super-elements × 100 time steps × 8 bytes/double), or about 400 KB, making it feasible to transfer and share via conventional networks. This facilitates the creation of large, shareable head FE simulation databases, accelerating the development and validation of brain injury research.

An analysis of the discrepancies between the reconstructed MPS data and the original THUMS data reveals that for most of the brain, the differences are small. This is likely because elements within the same anatomical region exhibit similar mechanical responses; thus, under various impact kinematics, they deform in a coordinated manner, resulting in similar MPS time-history curves that can be effectively represented by a single super-element. However, some regions exhibit larger errors, which we categorize into two types based on our observations.

1. **Contiguous High-Error Regions:** This is the primary source of reconstruction error for spatiotemporal MPS data. We hypothesize that in these regions, the mechanical responses of individual THUMS elements are highly variable, leading to significant differences in their MPS time-history curves. Given a fixed $K$, the representative capacity of a single super-element is limited, and it fails to capture the full diversity of responses within these heterogeneous regions. To test this hypothesis, we quantified the heterogeneity within each super-element by calculating the average Euclidean distance between the feature vector of each constituent element and the mean feature vector of the super-element. **Fig. 6a** illustrates the relationship between this heterogeneity metric and the mean peak MPS reconstruction RMSE for each super-element, averaged over the entire dataset. The Pearson correlation coefficient between these two metrics was 0.95 ($p < 0.0001$). This strong positive correlation supports our hypothesis that intra-super-element heterogeneity is the dominant systematic factor affecting the reconstruction fidelity of the templates.
2. **Sparsely Distributed, Isolated High-Error Elements:** This secondary source of error is associated with numerical artifacts in the original FE simulations. We posit that numerical artifacts, such as localized stress concentrations, can cause certain elements to exhibit abnormally high strain values. The reconstruction process, being an averaging operation, does not preserve these outliers, leading to a large discrepancy between the original and reconstructed values for these specific elements. To validate this, we selected a representative impact case containing significant numerical errors and analyzed the reconstruction error of its peak MPS. Since strain values greater than 1.0 are considered non-physiological for brain tissue, we classified all THUMS elements into two groups: "normal elements" (peak MPS < 1) and "numerical artifact elements" (peak MPS > 1). As shown in **Fig. 6b**, we found a statistically significant difference in the reconstruction errors between the two groups ($p < 0.0001$), with numerical artifact elements exhibiting much larger errors. This finding indicates that the observed discrete high-error points are consistent with the smoothing of localized peaks in the original FE outputs by the cluster-averaging operation. However, because no ground-truth strain field is available, we cannot determine whether this attenuation constitutes a valid correction of the original FE results. More specifically, the clustering-template representation implicitly performs a form of spatial filtering: during the THUMS-to-template projection, the MPS value of each super-element is computed as the average response of its constituent THUMS elements. As a result, isolated, non-physiological strain spikes are attenuated during projection and are not reproduced during reconstruction unless they are supported by a consistent response pattern among elements within the same cluster.

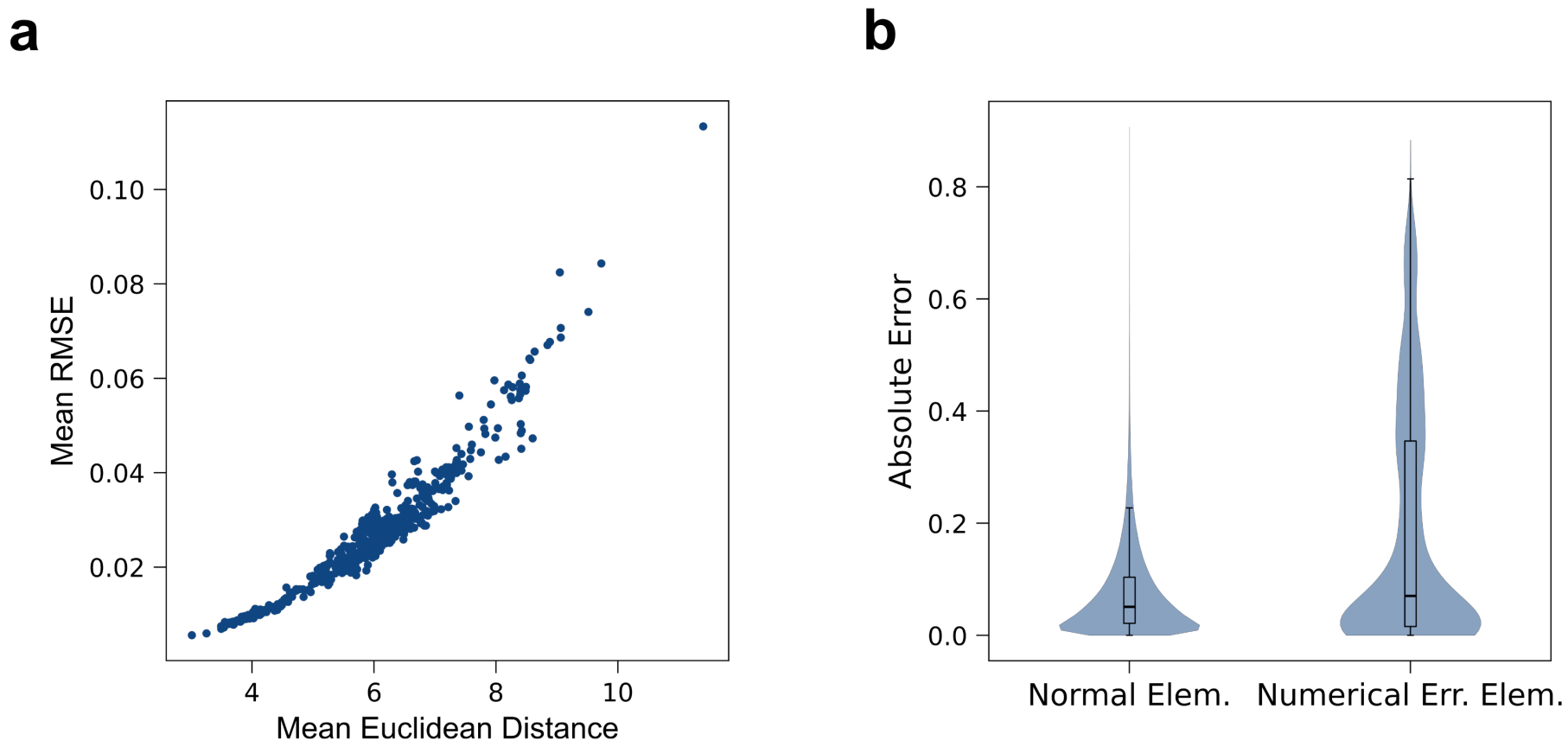

**Fig. 6 Analysis of reconstruction error sources. (a)** Correlation between reconstruction error and intra-super-element heterogeneity. The x-axis represents the average Euclidean distance between the feature vectors of constituent THUMS elements and the mean feature vector of the super-element. The y-axis represents the average peak MPS reconstruction RMSE for elements within that super-element, averaged over the entire dataset. **(b)** Distribution of peak MPS reconstruction error for normal vs. numerical error elements in a representative impact case (Impact 100) containing significant numerical artifacts.

While FE analysis is a cornerstone of modern biomechanics research for estimating tissue strain, the resulting spatiotemporal data are subject to inherent uncertainties and errors [68]. A 2018 study by Zhou et al. evaluated the well-validated KTH head model and found that even this advanced model produced local MPS predictions that differed notably from experimental values [69]. The comparison curves presented in their paper (Fig. 5 therein) show that the peak MPS values computed by the KTH model for various impacts typically deviated from cadaveric experimental data by 0.015 to 0.030. These model–experiment discrepancies provide practical numerical scale context while evaluating the reconstruction error. The mean peak-MPS reconstruction RMSE was 0.014 for K = 8192 and 0.027 for K = 512. Given the substantial reduction in data dimensionality, these values represent a reasonable approximation–compression tradeoff for applications focused on efficient storage, preprocessing, and large-scale analysis. Importantly, the reconstruction RMSE quantifies differences from the original THUMS outputs and does not necessarily correspond to an equivalent loss of physically meaningful strain information. Because the true physical strain field is unknown, we cannot determine whether the reconstruction-induced deviations partially attenuate discrepancies already present in the FE predictions or compound them.

The performance and appropriate granularity $K$ of the clustering templates are likely influenced by the modeling assumptions of the underlying FE model. For instance, the THUMS model used here features a smoothed brain surface. In contrast, models explicitly incorporating sulcal-gyral anatomy exhibit higher localized strain gradients at the cortical boundaries, which might necessitate a larger $K$ to maintain reconstruction fidelity. Moreover, while THUMS employs linear viscoelasticity, our clustering framework is inherently data-driven. It groups elements based on their observed biomechanical signatures (MPS time-history curves). Therefore, the framework remains applicable even when brain tissue enters a non-linear deformation regime or when hyperelastic constitutive laws are used, as it learns the coordinated deformation patterns (spatial co-variation) directly from the simulation data.

Our method for extracting a feature vector for each element—by sampling four time points per impact—was constrained by memory and computational resources and may not be the optimal approach. Although the four peak-centered time points selected during feature extraction provided sufficient information for high-fidelity reconstruction in this study, this feature-extraction strategy remains heuristic. Compared with using full MPS time histories, this peak-centered representation may underrepresent secondary peaks, prolonged low-amplitude responses, or impact-specific differences in response duration. Therefore, while the current feature extraction strategy provides a favorable balance between computational tractability and reconstruction fidelity, for impacts with substantially different temporal characteristics, adaptive time windows, full-time-history representations, or learned temporal embeddings may provide more robust features and should be investigated in future work[70]. Additionally, to safeguard downstream analyses from numerical instabilities, we implemented an upper-limit clipping of 1.0 on the original MPS data before template construction and error evaluation. While this effectively filters out non-physical numerical artifacts common in high-density FE simulations, it represents a simplified treatment of extreme strain fields. For future designs utilizing advanced solver configurations that are inherently free of numerical errors, such a clipping process could be bypassed to preserve full-range strain resolution.

Furthermore, the clustering templates are specific to the THUMS AM50 v4.02 model and cannot be directly transferred to other head FE models. While the proposed framework is model-agnostic in principle and can be applied to other FE models by constructing new model-specific templates, the templates reported here were learned from THUMS-derived MPS time-history features. Therefore, differences in mesh topology, anatomical representation, and constitutive formulation among FE head models may lead to different clustering structures. Although the framework is in principle extensible to a cross-model setting—by jointly clustering feature vectors from multiple models in a shared feature space—whether elements from models with different mesh topologies and constitutive formulations can be meaningfully co-clustered remains an open question. Additionally, as THUMS AM50 represents a 50th-percentile adult male, the learned clustering structure may not generalize to populations with distinct brain morphologies or to subject-specific geometries, which can affect the brain strain in head impacts[71–73]. Moreover, the reduced representation is best suited for data compression and ML preprocessing; tasks requiring full element-level mechanical fidelity should employ clustering templates with larger $K$ values or the original FE data. Finally, the reconstruction assessment in this study was in-sample, as the clustering templates were constructed and evaluated using the same dataset. Therefore, the reported reconstruction metrics quantify compression fidelity for the loading conditions represented during template construction and do not constitute an independent assessment of generalization to unseen impacts. Although the dataset included both synthetic and real-world head impacts, future work should construct templates using a training subset and evaluate their reconstruction performance on held-out impacts and independent impact datasets. For substantially different or out-of-distribution loading scenarios, a new template may need to be generated using simulation data representative of those conditions.

This study employed the standard Forgy method for initializing cluster centroids. Future research could explore alternative initialization strategies, such as K-means++, to improve centroid initialization and reduce sensitivity to random initialization. Nevertheless, in our high-dimensional, large-sample setting, we observed high consistency in the reconstruction errors of templates generated from different random initializations, suggesting that the impact of initialization may be limited. Finally, our reconstruction method—assigning the super-element's value to all its constituent elements—is a simple approach. More advanced reconstruction techniques, such as neural network-based upsampling methods, could potentially improve reconstruction accuracy and warrant future investigation. Additionally, future work should validate the utility of the clustering template representation in downstream machine learning tasks—such as concussion classification or TBI risk prediction—to determine whether the compressed representation retains information relevant to injury-related outcomes.

# Statements and Declarations

### Acknowledgements

This work was supported by the National Natural Science Foundation of China (12302414,12672417), Beijing Municipal Natural Science Foundation (7252287), Young Elite Scientists Sponsorship Program by The Chinese Society of Theoretical and Applied Mechanics (CSTAM 2023-XSC-HW3), the Fundamental Research Funds for the Central Universities Program (501RCQD2025110001, 501QYJC2024110002).

### Competing Interests

The authors declare that they have no competing interests.

### Author Contributions

**Junxi Liu:** Conceptualization, Methodology, Software, Formal analysis, Investigation, Writing – original draft, Visualization. **Xiaoyu Du, Jingyi Wu, Yang Wang, Chongmin Zhang, Yihan Zhang, Siyuan Lv:** Validation, Data curation, Writing – review & editing. **Li Wang:** Resources, Supervision, Writing – review & editing. **Xianghao Zhan, Zhou Zhou:** Methodology, Resources, Data curation, Writing – review & editing. **Yuzhe Liu:** Conceptualization, Resources, Supervision, Project administration, Funding acquisition, Writing – review & editing. All authors read and approved the final manuscript.

### Data Availability

The clustering templates are publicly available at: https://github.com/Thethumbtack/Clustering-Template-for-Brain-Strain.

Other raw and processed data that support the findings of this study are available from the corresponding author upon request.